\documentclass[%
 reprint,
 amsmath,amssymb,
 aps,
]{revtex4-2}

\usepackage{graphicx}
\usepackage{dcolumn}
\usepackage{bm}

\begin{document}

\preprint{APS/123-QED}

\title{Phase-dependent controllable field generation in a ring cavity resonator}

\author{Sanket Das}
\author{Tarak N. Dey}%
 \email{tarak.dey@iitg.ac.in}
\affiliation{Department of Physics, Indian Institute of Technology Guwahati, Guwahati 781039, Assam, India\\}

\date{\today}

\begin{abstract}
We investigate the control field phase-dependent output field transmission from a red detuned ring cavity optomechanical system. Our scheme displays a double transparency window in the presence of a strong control and a weak probe field. Additionally, we invoke an external mechanical pump to one of the movable mirrors to modulate its vibration. Complete control over the output field transmission can be achieved due to the combined effect of the amplitudes and the phases of the mechanical pump and the control field. Further, a tunable group delay of the probe pulse propagation can be obtained by the tailoring of the control field phase in the presence of a suitable mechanical drive. We further discuss the effect of control field phase on Stokes field generation via the four-wave mixing process. This scheme may find potential applications in weak signal sensing and all-optical communication purposes.
\end{abstract}

\maketitle

\section{Introduction}

The radiation pressure interaction between optics and mechanics opens up promising perspectives in the field of quantum optics and quantum information science.  Over the past decade, a tremendous effort has been made to understand various aspects of the radiation pressure interaction in gravitational-wave detection \cite{mcclelland2011advanced}, entanglement studies between two distant mechanical oscillators \cite{ockeloen2018stabilized,chakraborty2018entanglement}, photon and phonon blockade \cite{rabl2011photon,liao2013photon,xie2017phonon}. The conventional  optomechanical system has been extended to whispering-gallery-mode resonators \cite{schliesser2008high}, optomechanical crystals \cite{chan2012optimized,davanco2014si}, electromechanical circuits \cite{miao2012microelectromechanically}. It has also been realised that the radiation pressure force replicates many physical phenomena arising from the semi-classical behavior of the light-matter interaction. For instance, an optomechanical system shows a mechanical analog of electromagnetically induced transparency (EIT) \cite{harris1964two,harris1990nonlinear,boller1991observation,fleischhauer2005electromagnetically}. In EIT, a narrow spectral hole is formed in the probe absorption spectrum in presence of strong control field in an otherwise opaque atomic medium. The EIT has potential applications such as slow light \cite{hau1999light} and light storage\cite{liu2001observation}. The mechanical counterpart of EIT, commonly known as optomechanically induced transparency (OMIT) has been theoretically predicted \cite{agarwal2010electromagnetically,ma2014tunable,jing2015optomechanically,lu2017optomechanically} and experimentally observed both in the optical cavity \cite{safavi2011electromagnetically,weis2010optomechanically,chen2011slow,karuza2013optomechanically} as well as in the microwave cavity \cite{teufel2011circuit}. In an optomechanical system (OMS), the mean displacement of the movable mirror plays a crucial role to obtain a transparency window. The mirror mean displacement becomes zero for a quadratically coupled OMS. The fluctuations in the displacement of the movable mirror due to the interaction with the environment produces a narrow spectral window for the probe transmission \cite{huang2011electromagnetically,wu2020numerical,wang2019tunable,he2020normal}.
Contrary to EIT, electromagnetically induced absorption (EIA) phenomenon has been studied well  in a two-cavity optomechanical system as it mimics an $N$-type four-level atomic system \cite{qu2013phonon}. The controllable optical output field transmission, as well as absorption, has been investigated very recently in the presence of a weak periodic mechanical force \cite{xu2015controllable,suzuki2015nonlinear,lu2019selective,zhang2020mechanical,jiang2019phase,jiang2016phase,jia2015phase}.
The studies in this direction established that coherent interconversion between optical and mechanical excitations is possible in an optomechanical cavity \cite{groblacher2009observation,verhagen2012quantum,fiore2011storing}. Further, the mechanical excitations can be mapped to the optical field at different wavelengths referred to as the optical wavelength conversion \cite{tian2010optical,safavi2011proposal}.\\
In this work, we provide a theoretical study to facilitate controllable group delay of probe transmission in an optomechanical system. To achieve this, we use a phase-dependent strong control field and weak mechanical pump in a double OMIT configuration. For this purpose, a red detuned ring cavity is taken into account. 
We apply an external mechanical drive of phase $\phi_m$ to one of the movable mirrors. 
We find that the interference between the radiation pressure induced cavity fields and the mechanical pump induced sidebands leads to the enhancement of the output fields transmission while the mechanical drive is in phase with the applied probe field. However, a complete control on the output field transmission can be achieved by using the combined effect of the amplitude and the phase properties of the mechanical pump and the control field, respectively. Hence the control field phase $\phi_c$ invokes an extra tunability on output field transmission.  We also exhibit how the probe pulse propagation delay can be changed from slow to fast light by tailoring the phase of the control field. Further, we examine the effect of control phase on Stokes field generation via the four-wave mixing process.\\
The paper is organized as follows. In Sec.\ref{sec:Model}, the theoretical model for a ring cavity optomechanical system is described. This Sec.\ref{sec:Model} also contains the Heisenberg equation of motion to govern the temporal evolutions of the expectation values of each operator.  Section \ref{sec:Numerical Results} discusses the control field phase dependency on the output probe field transmission. Further, the group velocity of the optical probe pulse has been studied analytically and verified numerically in section \ref{sec:Group delay}. Section \ref{sec:Stokes field} addresses control field phase-dependent FWM field generation.  We draw our conclusions in section \ref{sec:Conclusions}.

\section{Theoretical model}
\label{sec:Model}
\begin{figure}[b!]
\centering
\includegraphics[width=\linewidth]{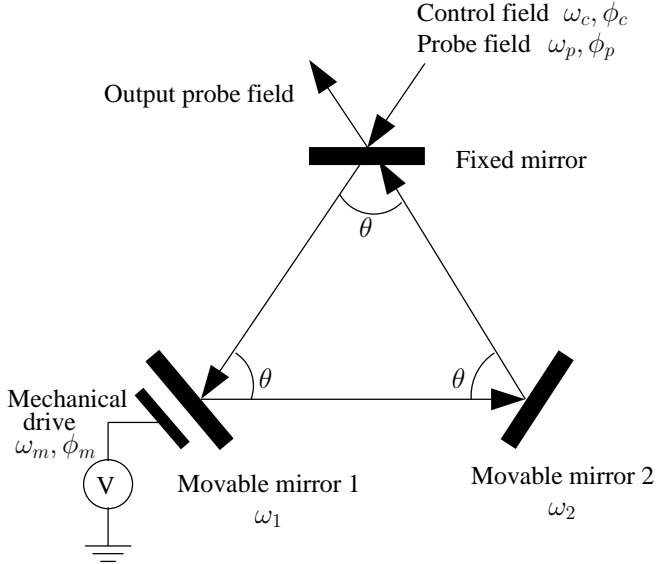}
\caption{\label{fig:model_diagram} The schematic diagram of a ring cavity resonator. A strong coupling field of frequency $\omega_c$ and a weak probe field of frequency $\omega_p$ are applied along with a weak mechanical drive of frequency $\omega_m$. The two movable mirrors have different resonant frequencies $\omega_1$ and $\omega_2$.}
\end{figure}

In this paper, we exploit the phase-sensitive behavior of the control and mechanical pump to manipulate the transmission and group velocity of the probe pulse through the ring cavity. For this purpose, we consider a ring cavity of length $L$, consisting of three mirrors as shown in Fig.\ref{fig:model_diagram}. Out of three mirrors, two are movable and perfectly reflecting mirrors, while the last one is fixed and partially transmitting. The frequency of the cavity field is $\omega_0$.  This cavity is strongly driven by a control field $\varepsilon_c$ with  frequency $\omega_c$ and phase $\phi_c$ together with a weak probe field $\varepsilon_p$ with frequency $\omega_p$ and phase $\phi_p$. In addition to that, we  apply a mechanical drive $\varepsilon_m$ with frequency $\omega_{m}$ and phase $\phi_m$ to one of the movable mirrors \cite{huang2014double,xu2015controllable}. The coupling between the cavity field and the movable mirrors can be achieved  through radiation pressure exerted by the photons in the cavity \cite{lebedev1901lebedev,nichols1903pressure}.
In our scheme, we treat the oscillation of the movable mirrors as quantum harmonic oscillators with resonance frequencies $\omega_i$, effective mass $m_i$, and mechanical damping $\gamma_i (i=1,2)$.  The damping of the mechanical oscillators arise from the interactions with the environment. We also assume that the mechanical frequencies $\omega_i$ are much larger than the cavity decay rate $ \kappa$ in order to produce well resolved sideband regime in OMIT \cite{weis2010optomechanically,agarwal2010electromagnetically,safavi2011electromagnetically}.
\newline
The total Hamiltonian of the model system can be expressed as
\begin{align}
\label{eq:Hamiltonian}
 H &=\hslash \omega_0 c^\dagger c +\frac{\hslash \omega_1}{2}\left(Q_1^2 + P_1^2\right)+\frac{\hslash \omega_2}{2}\left(Q_2^2 + P_2^2\right)\nonumber\\
 &+\hslash\left(g_1 Q_1 - g_2 Q_2\right)c^\dagger c \cos\frac{\theta}{2}+i\hslash\varepsilon_c \left(c^\dagger e^{-i\omega_c t-i\phi_c}- h.c\right)\nonumber\\
&+i\hslash\left(c^\dagger\varepsilon_p e^{-i\omega_p t-i\phi_p}- h.c\right)- 2 Q_1\varepsilon_m\hslash\cos(\omega_m t + \phi_m), 
\end{align}
where the first term corresponds to the free Hamiltonian of the
single-mode cavity with frequency $\omega_0$. The annihilation (creation) operator of the cavity field is denoted by $c(c^\dagger)$. The second and third term describes the total energy of two nanomechanical oscillators.
The fourth term illustrates the radiation pressure interaction between two movable mirrors and the cavity field. We introduced a relative minus sign between two radiation pressure terms which is consistent with the constant total cavity length, and the length between two mechanical mirrors are fixed during the oscillation.  The fifth and sixth terms come from the interaction between 
the cavity field and two input fields. The last term accounts for the external mechanical drive that is applied to one of the movable mirrors. In an electromechanical circuit, the time-dependent voltage source produces a current on the capacitor surface. The Coulomb interaction between these two conducting surface produces the periodic mechanical driving force \cite{zhang2012precision}. The dimensionless position and momentum quadratures of $j^{\textrm th}$ nano-mechanical oscillator are defined as $Q_j = q_j\sqrt{m_j \omega_j/\hslash}$ and $P_j= p_j\sqrt{1/m_j\omega_j\hslash}$ with commutation relation, $[Q_j,P_k] = i\delta_{jk}(j,k \in1,2)$. The amplitude of the control, probe, and the mechanical drive inside the ring cavity is given by
$\varepsilon_i=\sqrt{2\kappa P_i/\hslash\omega_i}$, $(i=c, p, m)$ with $P_i$,  being the respective power of fields and $\kappa$ the cavity decay rate. The Hamiltonian of the system in a rotating frame
with respect to the coupling field frequency $\omega_c$ is defined by $H_{rot}=RHR^\dagger+ i\hslash\left({\partial R}/{\partial t}\right)R^\dagger$,
where $R=e^{i\omega_c c^\dagger c t}$. By using  Baker-Campbell-Hausdorff formula
\begin{eqnarray}
 e^{\alpha A}Be^{-\alpha A}
=B+\alpha[A,B]+\frac{\alpha^2}{2!}[A,[A,B]]+...,\nonumber
\end{eqnarray}
 $H_{rot}$ can be cast into the following form
 \begin{align}
 H_{rot}&=\hslash \Delta_c c^\dagger c +\frac{\hslash \omega_1}{2}\left(Q_1^2 + P_1^2\right)+\frac{\hslash \omega_2}{2}\left(Q_2^2 + P_2^2\right)\nonumber\\
 &+\hslash\left(g_1 Q_1 - g_2 Q_2\right)c^\dagger c \cos\frac{\theta}{2}
 + i\hslash\varepsilon_c \left(c^\dagger e^{-i\phi_c}-h.c\right)\nonumber\\
 &+i\hslash\left(c^\dagger\varepsilon_p e^{-i\delta t-i\phi_p}- h.c\right)- 2 Q_1\varepsilon_m\hslash\cos(\omega_m t + \phi_m).
\end{align}
The detuning $\Delta_c=\omega_0-\omega_c$ 
and $\delta=\omega_p-\omega_c$. In this study, we treat the control and probe field as classical fields. We use the mean response of the system for investigating the transmitted output. The mean response of the system can be obtained by using the Heisenberg - Langevin equation as  $\langle\dot{\mathcal{O}}\rangle={i}\langle[H,\mathcal{O}]\rangle/\hbar+\langle N\rangle$,  under the rotating frame with frequency $\omega_c$. The $N$ stands for quantum fluctuation. Here we assume white noise as  a quantum fluctuation \cite{agarwal2010electromagnetically,huang2009normal}. By using Heisenberg - Langevin equation, we obtain a set of coupled differential equations as
\begin{align}
\label{eq:time-dependent}
 \langle\dot{c}\rangle&=-\left[\kappa+i\left(\Delta_c+\left(g_1 \langle{Q_1}\rangle - g_2 \langle{Q_2}\rangle\right)\cos\frac{\theta}{2}\right)\right]\langle{c}\rangle\nonumber\\
 &+\varepsilon_c e^{-i\phi_c}+\varepsilon_p e^{-i\phi_p}e^{-i\delta t},\nonumber \\
 \langle\ddot{Q_1}\rangle&=- \gamma_1\langle\dot{Q_1}\rangle - \omega_1^2\langle{Q_1}\rangle-\omega_1 g_1 \langle{c^\dagger}\rangle \langle{c}\rangle \cos\frac{\theta}{2}\nonumber\\
 &-2\varepsilon_m \omega_m\cos(\omega_m t+\phi_m),\nonumber \\
 \langle\ddot{Q_2}\rangle&=- \gamma_2\langle\dot{Q_2}\rangle - \omega_2^2 \langle{Q_2}\rangle+\omega_2 g_2 \langle{c^\dagger}\rangle \langle{c}\rangle\cos\frac{\theta}{2}.
 \end{align}
The radiation pressure force as well as the periodic mechanical force produce two sets of optical output fields at frequency $\omega_c \pm m\delta$, and frequency $\omega_c \pm n\omega_m$, respectively. Further, the resonant condition $\omega_m=\delta=\omega_1$ allows these two components of the output field to exhibit interference phenomena depending on the relative phase $\phi=\phi_p-\phi_m$ between the probe field and mechanical drive. In the strong driving field region, the mean values of any operator $\mathcal{O}(t)$ can be expressed as a sum of its steady-state value $\mathcal{O}_s$ and a small fluctuating time-dependent term $\tilde{\mathcal{O}}(t)$. The steady-state values of each operator are
 \begin{align*}
 Q_{1s} =-\frac{G_1}{\omega_1}c_s^*,~~Q_{2s} =\frac{G_2}{\omega_2}c_s^*,~~c_s =\frac{\varepsilon_c e^{-i\phi_c}}{\kappa+i \Delta'},
 \end{align*}
where $G_i=g_ic_s\cos{\theta}/{2}$ for $i=1,2$ are the effective optomechanical coupling rates and $\Delta'=\Delta_c+\left(g_1 Q_{1s}-g_2 Q_{2s}\right)\cos{\theta}/{2}$ is the effective detuning. Time dependent part of Eq. \ref{eq:time-dependent} reads
\begin{align}
\label{eq:coupled_equations}
\dot{\tilde{c}}+\left(\kappa+i\Delta'\right)\tilde{c}&=-iG_1\tilde{Q_1}+
iG_2\tilde{Q_2}+\varepsilon_pe^{-i\delta t}e^{-i\phi_p},\nonumber\\
 \ddot{\tilde{Q_1}}+\gamma_1\dot{\tilde{Q_1}}+\omega_1^2\tilde{Q_1}&=-\omega_1 G_1\tilde{c}^*-\omega_1 G_1^*\tilde{c}\nonumber\\
 &-2\varepsilon_m\omega_m\cos(\omega_mt+\phi_m),\nonumber\\
 \ddot{\tilde{Q_2}}+\gamma_2\dot{\tilde{Q_2}}+\omega_2^2\tilde{Q_2}&=\omega_2 G_2\tilde{c}^*+\omega_2 G_2^*\tilde{c}.
 \end{align}
We assume that the control field is much stronger than the probe field and mechanical drive. In this strong control field regime, we have adopted the following ansatz to solve the Eq. (\ref{eq:coupled_equations})
\begin{align}
\label{eq:ansatz}
 \tilde{c}=&c_{p+}e^{-i\delta t}+c_{p-}e^{i\delta t}+c_{m+}e^{-i\omega_m t}+c_{m-}e^{i\omega_m t},\nonumber\\
 \tilde{Q_1}=&Q_{1p}e^{-i\delta t}+Q_{1p}^*e^{i\delta t}+Q_{1m}e^{-i\omega_m t}+Q_{1m}^*e^{i\omega_m t},\nonumber\\
 \tilde{Q_2}=&Q_{2p}e^{-i\delta t}+Q_{2p}^*e^{i\delta t}+Q_{2m}e^{-i\omega_m t}+Q_{2m}^*e^{i\omega_m t}.
 \end{align}
We obtain
\begin{align}
\label{eq:optical sideband}
 c_{p+}(\delta)&=\frac{A(\delta)}{B(\delta)},~~
 c_{p-}(\delta)=\frac{F(\delta)}{H(\delta)}\\
 c_{m+}(\omega_m)&=\frac{D(\omega_m)}{E(\omega_m)},~~
 c_{m-}(\omega_m)=\frac{I(\omega_m)}{J(\omega_m)}.
 \end{align}
where
 \begin{equation}
\begin{aligned}
\label{eq:coefficients}
   A(\delta)&=\varepsilon_pe^{-i\phi_p}[-(i(|G_1|^2\omega_1(-i\gamma_2\delta-\delta^2+\omega_2^2)\\&+(-i\gamma_1\delta-\delta^2+\omega_1^2)
   ((\delta+\Delta'+i\kappa)(i\gamma_2\delta+\delta^2-\omega_2^2)\\
   &+|G_2|^2\omega_2)))],\nonumber\\
   B(\delta)&=((\delta-\Delta'+i\kappa)(\delta+\Delta'+i\kappa)(i\gamma_1\delta+\delta^2-\omega_1^2)\\
 &(i\gamma_2\delta+\delta^2-\omega_2^2)+G_2^2{G_1^*}^2\omega_1\omega_2\\
& +\omega_2G_2^*(2G_2\Delta'(-i\gamma_1\delta-\delta^2+\omega_1^2)
 +G_1^2G_2^*\omega_1)\\
 &-2|G_1|^2\omega_1(\Delta'(i\gamma_2\delta+\delta^2-\omega_2^2)+|G_2|^2\omega_2)),\nonumber\\
 D(\omega_m)&=\omega_1\varepsilon_me^{-i\phi_m}[(G_1(\Delta'+i\kappa+\omega_m)(i\gamma_2\omega_m+\omega_m^2-\omega_2^2)\\
    &+G_2\omega_2(-G_2G_1^*+G_1G_2^*))],\nonumber\\
  E(\omega_m)&=-({\Delta'}^2+(\kappa-i\omega_m)^2)(i\gamma_1\omega_m+\omega_m^2-\omega_1^2)\\
  &(i\gamma_2\omega_m+\omega_m^2-\omega_2^2)+(\omega_2\omega_1)(G_1^2{G_2^*}^2+G_2^2{G_1^*}^2)\\
  &+|G_1|^2(((\Delta'+i\kappa+\omega_m)(-i\gamma_2\omega_m-\omega_m^2+\omega_2^2)\\
  &-|G_2|^2\omega_2)\omega_1
    +\omega_1((\Delta'-i\kappa-\omega_m)(-i\gamma_2\omega_m-\omega_m^2+\omega_2^2)\\&-|G_2|^2\omega_2))
    -2|G_2|^2(\gamma_1\omega_m-i(\omega_m^2-\omega_1^2))\\
    &(\kappa\omega_2+i(\Delta'+i\kappa)\omega_2),\nonumber\\
  F(\delta)&=\varepsilon_pe^{i\phi_p}[\delta(G_1^2\gamma_2\omega_1+G_2^2\gamma_1\omega_2)+i(G_2^2\omega_2(\delta-\omega_1)(\delta+\omega_1)\\
  &+G_1^2\omega_1(\delta-\omega_2)(\delta+\omega_2))],\nonumber\\
 H(\delta)&=-(\delta^2-(\Delta'-i\kappa)^2)(-i\gamma_1\delta+\delta^2-\omega_1^2)(-i\gamma_2\delta+\delta^2-\omega_2^2)\\
 &+G_2^2{G_1^*}^2\omega_1\omega_2+G_1^2{G_2^*}^2\omega_1\omega_2
 -|G_2|^2(-i\gamma_1\delta+\delta^2-\omega_1^2)\\
 &((\delta+\Delta'-i\kappa)\omega_2+(\delta-\Delta'+i\kappa)\omega_2)
            -|G_1|^2(2\gamma_2\delta(-\kappa\omega_1\\
            &+(-i\delta+\kappa)\omega_1)+2((\Delta'-i\kappa)\omega_1+(\delta-\Delta'+i\kappa)\omega_1)\\
            &(\delta-\omega_2)(\delta+\omega_2)
           +2|G_2|^2\omega_1\omega_2),\nonumber\\
    I(\omega_m)&=-\varepsilon_m\omega_1e^{i\phi_m}(G_2^2G_1^*\omega_2-G_1(|G_2|^2\omega_2+(\Delta'+i\kappa-\omega_m)\\
    &(-\omega_2^2-i\gamma_2\omega_m+\omega_m^2))),\nonumber\\
    J(\omega_m)&=-({\Delta'}^2+(\kappa+i\omega_m)^2)(-i\gamma_1\omega_m+\omega_m^2-\omega_1^2)\\
    &(-i\gamma_2\omega_m+\omega_m^2-\omega_2^2)
            +G_2^2{G_1^*}^2\omega_1\omega_2+\omega_2G_2^*(2G_2\Delta'\\
            &(i\gamma_1\omega_m-\omega_m^2+\omega_1^2)+G_1^2G_2^*\omega_1\omega_2)+2|G_1|^2\omega_1(\Delta'\\
            &(i\gamma_2\omega_m-\omega_m^2+\omega_2^2)-|G_2|^2\omega_2).\nonumber\\
 \end{aligned}
\end{equation}
Here $c_{p+}$ and $c_{p-}$ are the output fields having frequencies the same as incident probe field and the generated FWM field, respectively. The mechanical pump induces two sidebands whose amplitude are given by $c_{m+}$ and $c_{m-}$. The output field from the cavity is obtained by the cavity input-output relation,
\begin{equation}
\label{eq:input_output}
 \varepsilon_{out}=2\kappa\langle{c}\rangle-\varepsilon_c-\varepsilon_pe^{-i\delta t-i\phi_p}-\varepsilon_me^{-i\omega_m t-i\phi_m}.
\end{equation}
By substituting Eq. \ref{eq:ansatz} into Eq. \ref{eq:input_output}, we obtain the normalised amplitude of the output probe field as
\begin{equation}
\label{eq:probe_output}
 t_p=\frac{2\kappa c_{p+}-\varepsilon_pe^{-i\phi_p}}{\varepsilon_pe^{-i\phi_p}}.
\end{equation}
Consequently, the transmitted amplitude of output FWM field at frequency $2\omega_c-\omega_p$ can be obtained as
\begin{equation}
\label{eq:FWM_output}
 t_f=\frac{2\kappa c_{p-}}{\varepsilon_pe^{i\phi_p}}.
\end{equation}
In addition, the mechanical drive can generate the upper sideband $\omega_c+\omega_m$ and lower sideband $\omega_c-\omega_m$ that takes following normalised forms
\begin{equation}
\label{eq:upper_lower}
 t_u=\frac{2\kappa c_{m+}-\varepsilon_m e^{-i\phi_m}}{\varepsilon_m e^{-i\phi_m}},~~{\textrm {and}}~~t_l=\frac{2\kappa c_{m-}}{\varepsilon_m e^{i\phi_m}}.
\end{equation}
Since we have considered $\omega_m=\omega_1=\delta$, therefore generated frequency of upper and lower sideband is same as probe frequency $\omega_p$ and FWM frequency $2\omega_c-\omega_p$, respectively. Then the output field amplitude oscillating at the probe frequency $t_{pu}$ and the FWM frequency $t_{fl}$ can be obtained from Eq. \ref{eq:optical sideband} as
\begin{equation}
\label{eq:probe}
 t_{pu}={}t_p+\eta t_ue^{i\phi}
\end{equation}
and
\begin{equation}
\label{eq:fourwave}
 t_{fl}={}t_f+\eta t_le^{-i\phi},
\end{equation}
 where $\eta=\varepsilon_m/\varepsilon_p$ and $\phi=\phi_p-\phi_m$ and their intensities are given by $|t_{pu}|^2$ and $|t_{fl}|^2$, respectively.
In order to study the transmitted fields from a ring cavity resonator, we use parameters similar to the previous studies \cite{lin2010coherent,huang2014double}. The strong control field of wavelength $\lambda=775$ nm couples to two movable mirrors having resonant frequencies $\omega_1=2\pi\times56.98$ MHz and $\omega_2=2\pi\times46.62$ MHz. Further, we consider the mass of these two mirrors to be the same as $m_1=m_2=20$ ng, the single-photon optomechanical coupling strengths are $g_1= 2\pi\times12$ GHz nm$^{-1}$ $\times\sqrt{\hbar/m_1 \omega_1}$ and $g_2= 2\pi\times12$ GHz nm$^{-1}$ $\times\sqrt{\hbar/m_2 \omega_2}$, the cavity decay rate $\kappa=2\pi\times 15$ MHz, the mechanical damping rates are $\gamma_1=\gamma_2= 2\pi\times 4.1$ KHz, and the angle between two successive arms of the ring cavity is $\theta=\pi/3$.
\section{Numerical results of the controllable output probe field generation}
\label{sec:Numerical Results}
\begin{figure}[t!]
\centering
\includegraphics[scale=0.42]{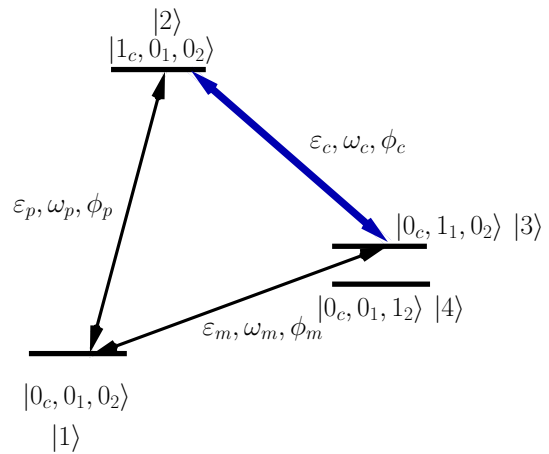}
\caption{\label{fig:level diagram}The level diagram of the ring cavity optomechanical system.}
\end{figure}

First, we study the analogy  between EIT in an atomic system and its mechanical counterpart. 
The mechanical system under consideration corresponds to the level diagram of Fig. \ref{fig:level diagram}, which constitutes an excited level $|2\rangle$, ground state $|1\rangle$ and two metastable states $|3\rangle$ and $|4\rangle$. In an atomic system,  the radiative decay rate $\gamma_{21}$ of the excited state $|2\rangle$ is much higher than the non-radiative decay rate $\gamma_{31}$ and $\gamma_{41}$ of metastable states $|3\rangle$ and $|4\rangle$. This is one of the criteria for formation of transparency window. In optomechanics, the cavity decay rate $\kappa$ and mechanical damping rate $\gamma_i(i\in1,2)$  play the same role as $\gamma_{21}$  and $\gamma_{i1} (i\in3,4)$ in an atomic system.\\ 

\begin{figure}[h!]
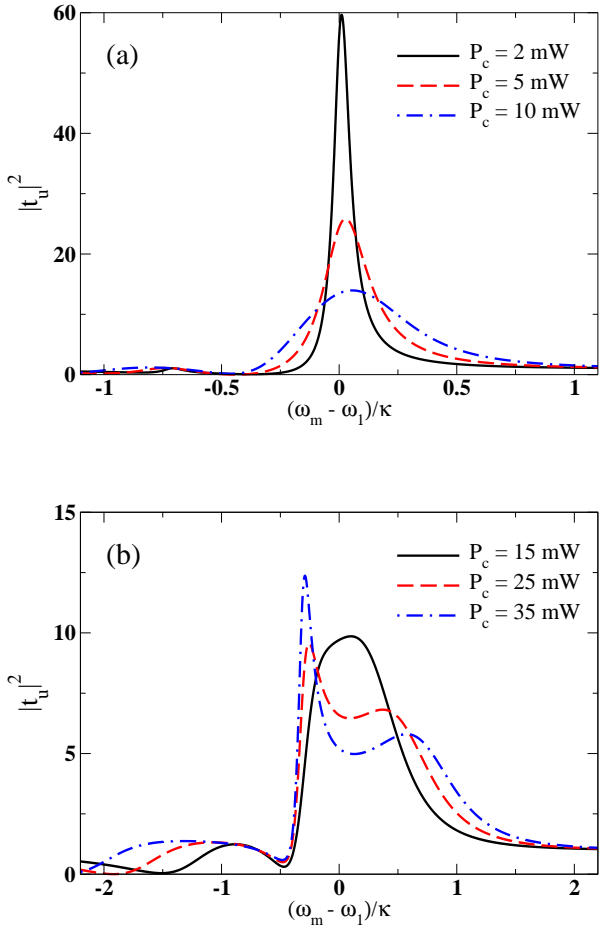
%
    \centering
   {\includegraphics[scale=0.26]{fig3a.eps} }\qquad\\
    \vspace{1cm}
    {\includegraphics[scale=0.26]{fig3b.eps} }%
    \caption{\label{fig:upper sideband}Transmission at the upper mechanical sideband is plotted against normalised mechanical drive detuning when the control field phase, $\phi_c=0$ and the mechanical drive is in-phase with applied probe field ($\phi_m=\phi_p$) for different control powers. The other parameters are $\omega_1=2\pi\times56.98$ MHz, $\omega_2=2\pi\times46.62$ MHz, $\gamma_1=\gamma_2= 2\pi\times 4.1$ KHz, $\kappa=2\pi\times 15$ MHz and the effective cavity detuning is $\Delta'=2\pi\times51.8$ MHz.}%
\end{figure}

\begin{figure}[t!]
\centering
\includegraphics[scale=0.26]{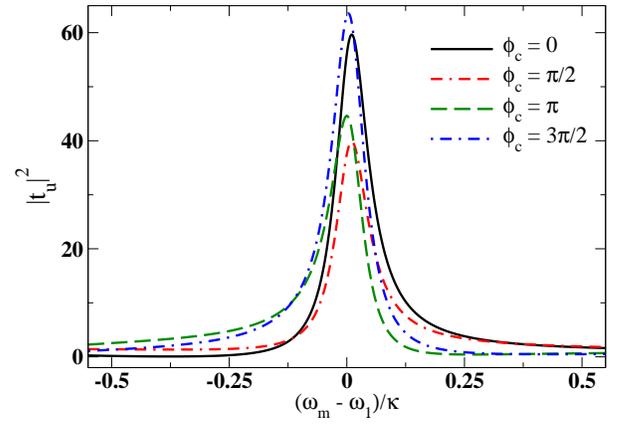}
\caption{\label{fig:phasedependent upper}The control phase $(\phi_c)$ dependency of the mechanical pump-induced upper sideband intensity is plotted against normalised mechanical drive detuning. The control power is kept at $2$ mW and the other parameters are the same as before.}
\end{figure}

\begin{figure}[h!]
\centering
{\includegraphics[scale=0.26]{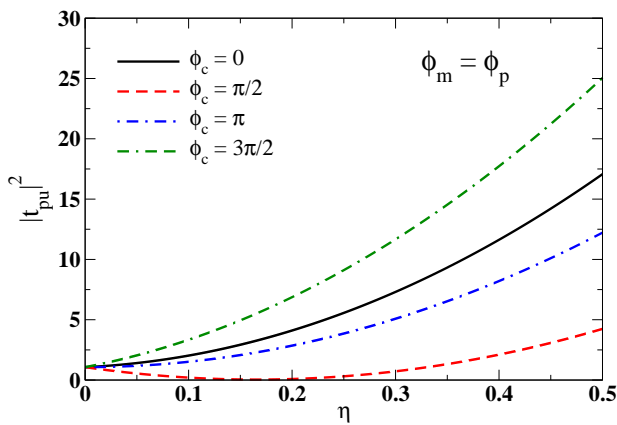}} \qquad
\caption{\label{fig:interference}The optical probe field transmission through the upper transparency window is plotted as a function of mechanical drive amplitude for different values of the control field phase. Here we consider the control field power to be $2$ mW.}
 \end{figure}

We now explore the weak mechanical pump induced upper mechanical sideband transmission at frequency $\omega_m+\omega_c$ given by Eq. \ref{eq:upper_lower}. In Fig. \ref{fig:upper sideband}, we plot the upper mechanical sideband intensity $|t_u|^2$ as a function of normalised mechanical drive detuning, $(\omega_m-\omega_1)/\kappa$. We assume that the mechanical drive is in-phase with the input probe field $\phi_m=\phi_p$ and the control phase $\phi_c$ is kept as zero. 
  The black solid curve in Fig. \ref{fig:upper sideband}(a) demonstrate the maximum peak of the transmission intensity at relatively low control powers. The sharp nature of the spectrum appears because, the coupling of the mechanical drive between $|1\rangle$ and $|3\rangle$ is dominant compared to the control field coupling between $|2\rangle$ and $|3\rangle$. Further increase in the control power results in a gradual decrease in the upper mechanical sideband intensity together with power broadening as exhibited by the red dashed line as well as blue dashed-dotted line in Fig. \ref{fig:upper sideband}(a). The normal mode splitting can be manifested in the upper mechanical sideband by increasing the control power as depicted in Fig. \ref{fig:upper sideband}(b).
The strong control field gives rise to an increase in the pumping rate between $|2\rangle$ and $|3\rangle$. Consequently, the optical pumping rate becomes more significant than the cavity decay rate $\kappa$. Therefore, the population spends a greater fraction of time in the excited state $|2\rangle$ that results in power broadening in the upper mechanical sideband as indicated by the black solid line in Fig. \ref{fig:upper sideband}(b). Further increasing the control power leads to the normal mode splitting as delineated by the red dashed line and blue dashed-dotted line in Fig. \ref{fig:upper sideband}(b). It is clear from these two curves that the position of one normal mode is almost unaffected, whereas the distance between the other two normal modes keeps on increasing with larger control power \cite{huang2014double}. As a result, the mechanical drive with frequency $\omega_m$ fails to couple resonantly with the mechanics $\omega_1$, and an asymmetric power spectrum arises as shown in Fig. \ref{fig:upper sideband}(b). Moreover, the graphical nature is determined by the roots of $E(\omega_m)$ \cite{huang2014double, agarwal2010electromagnetically}, which are in general complex. The real parts of the roots determine the spectral line peak positions and imaginary parts are associated with their widths. The pole structures can be found out by considering the control power to be $35$ mW with all other parameters remaining the same as mentioned earlier. The real parts of the roots of $E(\omega_m)$ provide the intensity maximum when normalised mechanical drive detuning, $(\omega_m-\omega_1)/\kappa$ values $-1.65, -0.32$ and $0.71$ are in well agreement with Fig. \ref{fig:upper sideband}(b).\\
Next, we explore the phase-sensitive behavior of the control field $\phi_c$ on the upper mechanical sideband transmission which is displayed in Fig. \ref{fig:phasedependent upper}. We have taken the pump power as $2$ mW and the other field phases to be equal {\it i.e.,} $\phi_m$=$\phi_p$. From Fig. \ref{fig:phasedependent upper}, we find that reduction of the upper sideband transmission is possible by changing the phase of the control field. However, the complete suppression of the upper sideband transmission is prohibited due to the dominance of coherent coupling of the mechanics over the radiation pressure coupling. Much clearer evidence of the role of the control field phase on probe transmission can be found by studying Eq. \ref{eq:probe}. Equation \ref{eq:probe} leads to $|t_{pu}|^2=|t_p|^2 + |\eta t_u|^2+\eta t_p t_u^*e^{i(\phi_m-\phi_p)}+ \eta t_p^* t_u e^{-i(\phi_m-\phi_p)}$ which corresponds to the phase-dependent transmission of the output probe field. The cross-terms give rise to the interference effect between cavity generated probe field and the upper mechanical sideband.  The interference terms can be survived only for the non-zero value of $t_u$ at $\phi_m$=$\phi_p$. Hence, the control field phase $\phi_c$ assisted  $t_u$ manipulation can allow the controlling the output probe transmission at a fixed value of $\phi$ ($\it{e.g.}$ $\phi=0$). The output probe field intensity can be further tamed by adjusting the phase difference between probe and mechanical components $\it{i.e.,}$ $\phi=\phi_m-\phi_p$.
  \begin{figure}[b!]
 \centering
\includegraphics[scale=0.25]{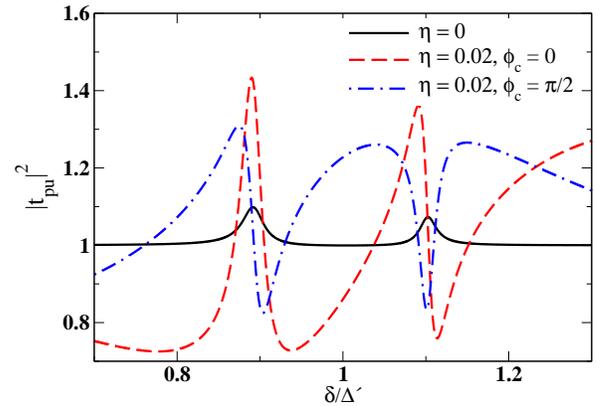} 
\caption{\label{fig:controllable probe}The control field phase-dependent transmission of the output probe field is plotted as a function of detuning between the control and the probe field when the mechanical drive is in-phase with the input probe field. All other parameters are the same as before.}
\end{figure}
In Fig. \ref{fig:controllable probe}, the solid black curve depicts two transparency windows for the output probe field which are created in the absence of mechanical drive. These two transparency windows are located around $\delta\approx\omega_1$ and $\omega_2$ with slightly unequal amplitudes. The unequal amplitudes emerge due to the difference in radiation pressure force experienced by the two nearly degenerate oscillators. Fig. \ref{fig:controllable probe} also confirms that the enhancement as well as suppression of the probe  transmission is possible in the presence of a mechanical pump. Note that the mechanical pump is in-phase with the input probe field. The values of control phase $\phi_c$  for the red dashed and blue dot-dashed curves of Fig. \ref{fig:controllable probe}, are $0$ and $\pi/2$, respectively.\\
We next focus on the probe transmission at the upper transparency window placed at $\delta=\omega_1$. In Fig. \ref{fig:interference}, we have plotted probe transmission as a function of normalised mechanical drive amplitude $\eta$ for different values of $\phi_c$. The constructive interference between the cavity-generated probe field and the mechanical pump-induced upper sideband, substantially enhances the output probe intensity for $\phi_c= 0,\pi,$ and $3\pi/2$. Nonetheless, the destructive interference between these two components forces the probe transmission to diminish, as seen from the red dashed curve at $\eta=0.18$.\\
 \subsection{Group delay}
 \label{sec:Group delay}
\begin{figure}[h!]
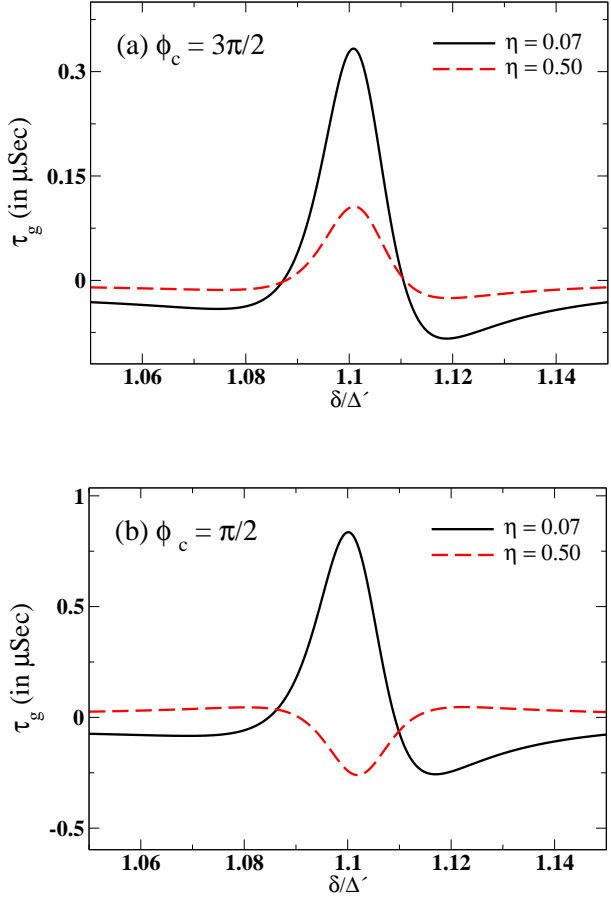

\centering
{\includegraphics[scale=0.26]{fig7a.eps}}\qquad\\
\vspace{1cm}
{\includegraphics[scale=0.26]{fig7b.eps}}\qquad
\caption{\label{fig:delay}Time delay of the probe pulse for different mechanical drive amplitudes have been plotted against the probe detuning $\delta/\Delta'$ while the control power is $2$mW. The mechanical drive is considered to be in-phase with the probe field. All other parameters are taken as the same as Fig. 3.  }
 \end{figure}

\begin{figure}[h!]
\centering
\includegraphics[width=\linewidth]{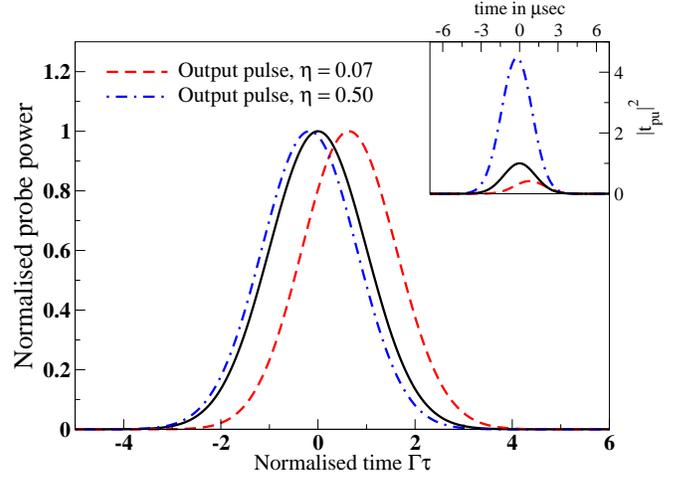}
\caption{\label{fig:pulse propagation}The normalised probe pulse intensity with respect to its peak value is plotted against the normalised time $\Gamma \tau$. The solid black curve represents the input probe pulse propagating at speed $c$ through the ring cavity. The red dashed and the blue dot-dashed curves represent the output probe pulse in the presence of mechanical pump with $\eta=0.07$ and $0.5$, respectively. Variation of the relative intensity of the output probe pulse is shown in the inset.}
\end{figure}

 Superposition of monochromatic waves centred around a carrier frequency $(\omega_s)$ can give rise to a light pulse with finite width. The envelope of the light pulse moves with a group velocity inside a dispersive medium. The existence of a transparency window is mandatory for pulse propagation through the otherwise opaque medium. To construct analytical expression for the group delay, we consider the envelope of the optical probe pulse as
\begin{equation}
 f(t_0)=\int_{-\infty}^{\infty}\tilde{f}(\omega)e^{-i\omega t_0}d\omega,\nonumber\\
\end{equation}
where $\tilde f(\omega)$ corresponds to the envelope function in the frequency domain. Accordingly, the reflected output probe pulse can be expressed as
\begin{align}
 \label{eq:definition}
 f^R(t_0)=&\int_{-\infty}^{\infty}t_{pu}(\omega)\tilde{f}(\omega)e^{-i\omega t_0}d\omega,\nonumber\\  
 =&e^{-i\omega_s t_0}\int_{-\infty}^{\infty}t_{pu}(\omega_s+\delta)\tilde{f}(\omega_s+\delta)e^{-i\delta t_0}d\delta.
\end{align}
Now we can expand $t_{pu}(\omega_s+\delta)$ in the vicinity of $\omega_s$ by Taylor series expansion and keeping the terms upto first order in $\delta$
\begin{equation}
  \label{eq:taylorexpansion}
  \begin{aligned}
    t_{pu}(\omega_s+\delta) &\approx t_{pu}(\omega_s)+\delta\frac{dt_{pu}}{d\omega}\bigg|_{\omega_s}.\\       
  \end{aligned}
\end{equation}
On substituting Eq. \ref{eq:taylorexpansion} into Eq. \ref{eq:definition}, we find
 \begin{align}
 \label{eq:intermediate}
 f^R(t_0)=&e^{-i\omega_s t_0}\int_{-\infty}^{\infty}t_{pu}(\omega_s)e^{-i\delta\left(t_0+\frac{i}{t_{pu}(\omega_s)}\frac{dt_{pu}}{d\omega}|_{\omega_s}\right)}\tilde{f}(\omega_s+\delta)d\delta,\nonumber\\
 =&e^{-i\omega_s t_0}\int_{-\infty}^{\infty}t_{pu}(\omega_s)e^{-i\delta\left(t_0-\tau_{g}\right)}\tilde{f}(\omega_s+\delta)d\delta\nonumber\\
 =&t_{pu}(\omega_s)e^{-i\omega_s\tau_{g}}f(t_0-\tau_{g}),
 \end{align}
where time delay is defined as \cite{safavi2011electromagnetically}
\begin{equation}
  \label{eq:timedelay}
  \begin{aligned}
    \tau_g= {\textrm Re}\left[\frac{-i}{t_{pu}(\omega_s)}\left(\frac{dt_{pu}}{d\omega}\right)\bigg|_{\omega_s}\right].\\\       
  \end{aligned}
\end{equation}
We examine the effect of a weak mechanical pump on the probe pulse propagation delay as depicted in Fig. \ref{fig:delay}. Here, we focus on the upper transparency window centered at $\delta\approx\omega_1$ and the control field intensity $2$ mW and phase $3\pi/2$, respectively.
Additionally, the input probe pulse is in phase with the external mechanical drive. The black solid curve of Fig. \ref{fig:delay}(a) corresponds to the probe pulse propagation delay $\tau_g=0.33$ $\mu$sec for $\eta=0.07$. Further increase in the mechanical pump amplitude causes a gradual decrease in the probe pulse propagation delay. Even then,  it displayed  slow light phenomena as confirmed from the red dashed curve in Fig. \ref{fig:delay}(a).
In Fig. \ref{fig:delay}(b), we change the control phase to $\pi/2$ and keep all other parameters the same as mentioned before. For a mechanical pump with amplitude $\eta=0.07$, we observe the delay of the output probe pulse as $0.83$ $\mu$sec. It is also evident from the red dashed curve in Fig. \ref{fig:delay}(b) that a relatively stronger mechanical pump can switch the light propagation velocity from slow to fast in the presence of control phase $\phi_c=\pi/2$. The probe pulse advancement corresponding to the red dashed curve is $0.27$ $\mu$sec for $\eta=0.5$. \\
To verify the above result, we consider the carrier frequency of the Gaussian probe pulse centered at the upper transparency window {\it i.e.,} $\delta=\omega_1$. The envelope of the pulse can be written as
\begin{equation}
\label{eq:pulsedefinition}
\tilde{f}(\omega)=\frac{\varepsilon_p}{\sqrt{\pi\Gamma^2}}e^{-\frac{(\omega-\omega_s)^2}{\Gamma^2}},\nonumber
 \end{equation}
where $\Gamma$ is the spectral width of the pulse. We keep in mind that the spectrum width $\Gamma$ must be well-contained in the transparency window for a distortionless probe pulse propagation. Successively numerically integrating Eq. \ref{eq:definition} bring out the temporal profile of the output probe pulse. The pulse parameters such as  gain or absorption and group velocity can be obtained by examining the nature of the transmission coefficient $t_{pu}$. For numerical integration, we have used the control power $2$ mW with phase $\pi/2$. The temporal intensity profile of the input probe pulse with $\Gamma$ as $2\pi\times129.5$ kHz is presented by the solid black curve of Fig. \ref{fig:pulse propagation}.
The utilisation of the weak mechanical driving field $\eta=0.07$ is visible from the red dashed curve of Fig. \ref{fig:pulse propagation} that exhibits the slow pulse propagation. The group delay $0.8$ $\mu$sec is measured from the peak separation between the input pulse and the output pulse. The sensitive nature of the group delay on mechanical pump amplitude can be seen in the blue dot-dashed curve of Fig. \ref{fig:pulse propagation}.  We estimate an advancement of the probe pulse around $0.24$ $\mu$sec. The physics behind the formation of slow and fast probe pulse turns out to be a reasonable suppression or enhancement of amplitudes due to interference of phases among the constituents waves. It can be understood from Eq. \ref{eq:probe}, which can also be expressed as $|t_{pu}|^2=|t_p|^2 + |\eta t_u|^2+\eta t_p t_u^*e^{i(\phi_m-\phi_p)}+ \eta t_p^* t_u e^{-i(\phi_m-\phi_p)}$. Here the cross-terms signify the destructive interference between cavity-generated probe pulse and the mechanical pump-induced upper sideband. For low mechanical drive amplitude, the dominance of the destructive interference  over the upper sideband intensity results in a reasonable discretion in the pulse output.
However, a relatively stronger mechanical drive amplitude provides the dominance of the upper sideband intensity over the destructive interference leading to an enhancement in output probe power. As depicted in the red dashed and blue dash-dotted curve of the inset of Fig. \ref{fig:pulse propagation}, the relative power of the output field with respect to input intensity are $0.42$ and $4.46$ for $\eta=0.07$ and $0.5$, respectively. Interestingly, the temporal width of the probe pulse is almost unaltered during the propagation through the ring cavity optomechanical system. This numerical result is in very good agreement with our analytical results for the probe pulse propagation delay as depicted in Fig. \ref{fig:delay}(b). Finally the mechanical coupling intensity and control phase play key roles in changing the probe pulse velocity from slow to fast light.
 \section{Controllable Stokes field generation}
 \label{sec:Stokes field}
In the context of optomechanics, the Stokes field generation through the nonlinear four-wave mixing process has been investigated quite extensively in previous literature \cite{huang2010normal,jiang2012controllable,he2020normal}. Further studies \cite{xu2015controllable} illustrate that the phase difference between the applied mechanical pump and the input probe field ($\it{e.g.}$ $\phi=\phi_m-\phi_p$) leads to the controllable Stokes field generation ($2\omega_c-\omega_p$) while the resonance condition ($\delta=\omega_m=\omega_1$) is well satisfied. To the best of our knowledge, the control field phase $\phi_c$ dependency of the Stokes field generation in a mechanically driven optomechanical system has not been studied till date. 
\begin{figure}[h!]
\centering
\includegraphics[scale=0.24]{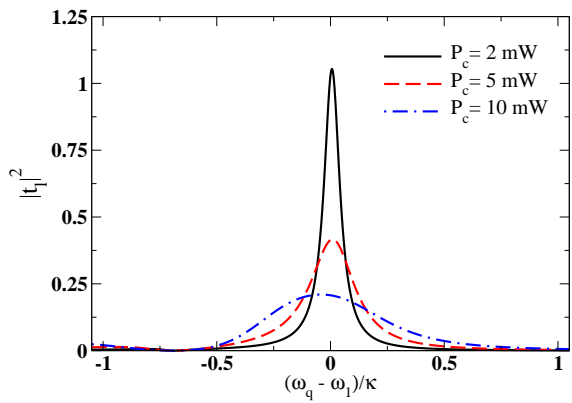}
\caption{\label{fig:lowersideband}The lower sideband transmission is plotted against normalised mechanical drive detuning for three different control powers $P_c= 2, 5$ and $10$ mW. All other parameters are same as in Fig. 3. }
\end{figure}
\begin{figure}[t!]
\begin{center}
\includegraphics[scale=0.24]{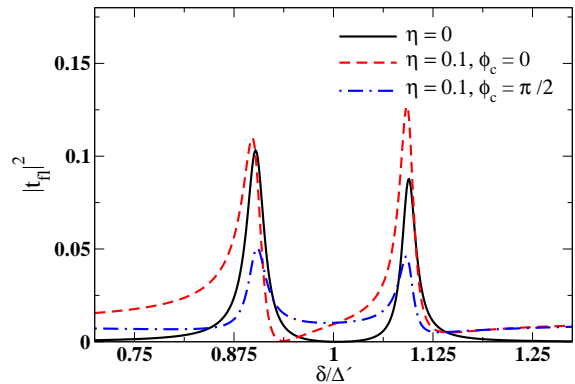}\qquad
\caption{\label{fig:cont_stokes}The control field phase-dependent Stokes field generation is plotted as a function of detuning between the control and the probe field. The mechanical drive is in-phase with the input probe field. The control power is taken as $2$ mW. All other parameters are same as in Fig. 3.}
\end{center}
 \end{figure}
 \\
 To study the Stokes field generation in an optomechanical system, we consider Eq. \ref{eq:fourwave} which can be expressed as $|t_{fl}|^2=|t_f|^2 + |\eta t_l|^2+\eta t_l^*t_fe^{i\left(\phi_m-\phi_p\right)}+\eta t_f^* t_le^{-i\left(\phi_m-\phi_p\right)}$. The first term corresponds to the radiation pressure induced Stokes field generation. The second term represents the external mechanical pump induced lower mechanical sideband generation. Whereas, the cross-terms emulate the effect of interference between these two components. In Fig. \ref{fig:lowersideband}, the lower mechanical sideband intensity ($|t_l|^2$) has been displayed as a function of normalised mechanical drive detuning, $(\omega_m-\omega_1)/\kappa$.
The black solid curve of Fig. \ref{fig:lowersideband} demonstrates a significant lower mechanical sideband generation as the mechanical coupling dominates over the weak radiation pressure coupling.
Further increase in the control power results in a gradual decrease in the lower sideband intensity along with power broadening as presented by the red dashed and blue dash-dotted lines of Fig. \ref{fig:lowersideband}. It is also evident from Fig. \ref{fig:lowersideband} that the control field phase dependency is absent due to the absolute square of lower sideband amplitudes. The importance of the control field phase becomes distinct after the expansion of $|t_{fl}|^2$.  The phase-induced interference effects that arise from cross terms can persist when the lower mechanical sideband amplitude is complex at $\phi=0$. The complex nature of the lower mechanical sideband comes from the phase dependence of the control field, $\phi_c$. Hence, the control field phase $\phi_c$ can be used to manipulate the interference terms that allow controlling the Stokes field generation even for a fixed value of $\phi$ ({\it e.g.} $\phi=0$) as shown in Fig.\ref{fig:cont_stokes}.
\begin{figure}[b!]
{\includegraphics[scale=0.24]{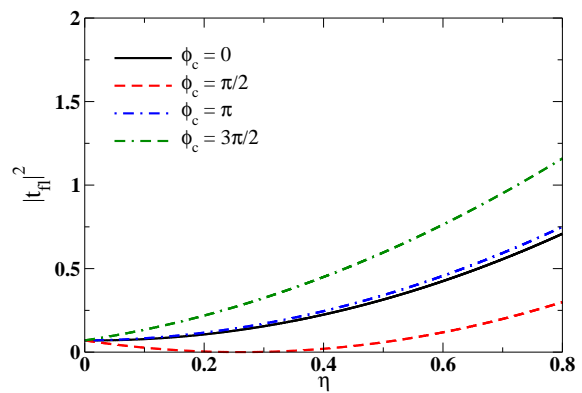}} \qquad
\caption{\label{fig:int_stokes}The Stokes field generation through the upper transparency window is plotted as a function of mechanical drive amplitude for different values of the control phase. The external mechanical pump is considered to be in-phase with the applied probe field ($\phi_m=\phi_p$). Here we consider the control field power to be $2$ mW.}
 \end{figure}
In the absence of the mechanical drive, the radiation pressure interaction with two nearly degenerate harmonic oscillators produces the Stokes field centered around $\delta\approx\omega_1$ and $\omega_2$ with slightly different amplitudes as shown by the black solid line. The enhancement as well as suppression of the Stokes field generation can be achieved by the application of a weak mechanical pump at different values of control phase $\phi_c$. The values of $\phi_c$ for the red-dashed and blue dot-dashed lines of  Fig. \ref{fig:cont_stokes} are $0$ and $\pi/2$, respectively. \\
We next concentrate on the Stokes field generation around the upper transparency window ($\delta=\omega_1$). In Fig. \ref{fig:int_stokes}, we have plotted $|t_{fl}|^2$ as a function of normalised mechanical drive amplitude $\eta$ for different control phases $\phi_c$. The constructive interference between the cavity-generated Stokes field and the mechanical pump-induced lower sideband enhances the output Stokes field intensity for the control phases $0,\pi$, and $3\pi/2$. The destructive interference between these two components subdues the output Stokes field generation for the normalised mechanical drive amplitude $\eta\approx0.3$ as seen from the red dashed curve of Fig. \ref{fig:int_stokes}. Hence, the phase-dependent control field is the main key behind the enhancement or complete suppression of Stokes field generation in a ring cavity.
\section{Conclusion}
\label{sec:Conclusions}
In conclusion, we have carried out theoretical investigations on control field phase assisted tunable group delay of the probe transmission as well as efficient FWM field generation in a ring cavity optomechanical system.
The optomechanical system consists of a red-detuned ring cavity that is an assembly of two nearly degenerate movable mirrors. A double EIT window is formed by the application of strong control and weak probe fields provided the two-photon resonance condition is satisfied. We apply an external mechanical drive of phase $\phi_m$ to modulate the vibration of one of the movable mirrors.
In the presence of $\phi_m$, we found that the strength of the output probe field can be enhanced due to the interference between the cavity-generated probe field and the mechanical drive-induced upper sideband. Moreover, the control field phase-sensitive behavior of the interference provides an extra tunability to control the output probe power. We demonstrate that a tunable group delay of the probe pulse in the course of propagation through the upper transparency window can be achieved by using a suitable amplitude of the mechanical pump and the phase of the control field while the relative phase between the probe field and the mechanical drive is kept fixed. To verify this claim, we also study the propagation of a Gaussian probe pulse that is well-contained in the upper transparency window.  The presence of the external mechanical pump amplitude along with suitable control field phases can switch the group delay of the probe pulse from slow to fast light. Further, the control field phase can enhance and surpasses the Stokes field generation via FWM. Our work may find potential application for storage and retrieval of the optical signals with the aid of the mechanical oscillator which provides a much longer lifetime than the atomic medium.
\nocite{*}

\bibliography{reference}
\end{document}